\documentclass[aps,prl,twocolumn,superscriptaddress,longbibliography]{revtex4-2}
\usepackage{amsmath,amssymb}
\usepackage[pdftex]{hyperref,graphicx}
\hypersetup{colorlinks = true, urlcolor = blue, linkcolor = blue, citecolor = blue}
\usepackage{physics}
\usepackage{xcolor}
\usepackage{bm}
\usepackage{pdfpages}

\newcommand{\comment}[1]{{}}

\makeatletter
\AtBeginDocument{\let\LS@rot\@undefined}
\makeatother

\begin{document}
\title{Anderson localization crossover in 2D Si systems: The past and the present}
\author{Seongjin Ahn}
\affiliation{Condensed Matter Theory Center and Joint Quantum Institute, Department of Physics, University of Maryland, College Park, Maryland 20742, USA}
\author{Sankar Das Sarma}
\affiliation{Condensed Matter Theory Center and Joint Quantum Institute, Department of Physics, University of Maryland, College Park, Maryland 20742, USA}

\begin{abstract}
Using Ioffe-Regel-Mott (IRM) criterion for strong localization crossover in disordered doped 2D electron systems, we theoretically study the relationships among the three key experimentally determined localization quantities: critical density ($n_\mathrm{c}$), critical resistance ($\rho_\mathrm{c}$), and sample quality defined by the effective impurity density (as experimentally diagnosed by the sample mobility, $\mu_\mathrm{m}$, at densities much higher than critical densities). Our results unify experimental results for 2D metal-insulator transitions (MIT) in Si systems over a 50-year period (1970-2020), showing that $n_\mathrm{c}$ ($\rho_\mathrm{c}$) decrease (increase) with increasing sample quality, explaining why the early experiments in the 1970s, using low-quality samples ($\mu_\mathrm{m} \sim 10^3 \mathrm{cm}^2/Vs$) reported strong localization crossover at $n_c \sim 10^{12} \mathrm{cm}^{-2}$ with $\rho_c \sim 10^3\Omega$ whereas recent experiments (after 1995), using high-quality samples ($\mu_\mathrm{m} >10^4 \mathrm{cm}^2/Vs$), report $n_c \sim 10^{11} \mathrm{cm}^{-2}$ with $\rho_c>10^4\Omega$.  Our theory establishes the 2D MIT to be primarily a screened Coulomb disorder-driven strong localization crossover phenomenon, which happens at different sample-dependent critical density and critical resistance, thus unifying Si 2D MIT phenomena over a 50-year period.
\end{abstract}

\maketitle
{\em Introduction.}---  
Anderson established in 1958 that a strongly disordered system may localize electrons due to the destructive interference of the electron waves scattering from random disorder \cite{andersonAbsenceDiffusionCertain1958}. This is one of the cornerstones of modern condensed matter physics as the localization transition is universal in itinerant electron (or bosonic) systems with increasing disorder (or deceasing carrier density) with the system undergoing a MIT from a metal to an insulator driven by disorder-induced  quantum interference leading to electrons localization.  It was later realized in \cite{abrahamsScalingTheoryLocalization1979} and \cite{wegnerElectronsDisorderedSystems1976, wegnerMobilityEdgeProblem1979} that in 2D systems, the localization transition is actually a crossover from an essentially unobservable (except perhaps at extremely low temperatures) logarithmically weakly localized  effective metal (for any finite system) to an exponentially localized  Anderson insulator at some critical density, but this crossover can be very sharp, making the 2D localization crossover appear operationally almost identical to the corresponding  Anderson localization in 3D disordered systems where the localization transition is a thermodynamic quantum phase transition. We ignore the subtle and very small weak localization corrections in the current work and focus on the experimentally reported density-tuned crossover  from the higher-density effective metallic phase to a lower-density effective strongly localized phase \cite{bishopNonmetallicConductionElectron1980a, urenLogarithmicCorrectionsTwodimensional1981, vankeulsLocalizationScalingRelation1997, luTerminationTwoDimensionalMetallic2011, minkovGiantSuppressionDrude2007, minkovQuantumCorrectionsConductivity2002, dassarmaSignaturesLocalizationEffective2014, altshulerWeaklocalizationTypeDescription2000a, altshulerMetalinsulatorTransition2D2001, prusCoolingElectronsSilicon2001, dassarmaSignaturesLocalizationEffective2014, brunthalerWeakLocalization2D1999}. This crossover is known as 2D MIT.

The current work is on the theory of 2D localization crossover in 2D Si systems, which has been extensively studied experimentally since the early 1970s \cite{andoElectronicPropertiesTwodimensional1982}. The great advantage of 2D Si systems (e.g., Si MOSFET) is that the carrier density can be easily tuned in a single sample over a large range just by changing a gate voltage, thus exploring the whole conductor-to-insulator regime from deep in the metallic phase with high conductivity at high density to the strongly localized insulator at low densities where the resistivity basically diverges exponentially at low temperatures. Since metallic conductors and localized insulators are operationally distinguished by the temperature dependence of their low-temperature resistivity, with the metallic (insulating) resistivity increasing as a power-law or remaining approximately constant (decreasing exponentially) with increasing temperature, it is easy to experimentally determine the critical crossover density $n_\mathrm{c}$ (and the associated critical resistance $\rho_\mathrm{c}$) by identifying the density at which the temperature dependence first manifests an exponentially increasing behavior as temperature is lowered. The ability to tune the carrier density in a single sample (i.e., with a fixed quenched disorder) to obtain $n_\mathrm{c}$ and $\rho_\mathrm{c}$ has made 2D Si MOSFETs the ideal system to study Anderson localization for almost 50 years, ever since experimental techniques were developed to do transport measurements of Si MOSFETs at cryogenic temperatures. It is therefore no surprise that the very first ($\sim1975$) disorder-driven and density-tuned Anderson localization from a conducting metallic phase to an activated transport exponentially localized insulating  phase with decreasing carrier density was first reported in 2D MOSFETs \cite{tsuiLocalizationMinimumMetallic1975, tsuiMottAndersonLocalizationTwoDimensional1974, pepperVariablerangeHoppingSilicon1974, mboungaHydrogenicAtomMultiphonon1974, arnoldDisorderInducedCarrier1974, pollittAndersonTransitionSilicon1976, pepperSpatialExtentLocalized1974}. These early experiments of the 1970s used relatively dirty samples with $\mu_\mathrm{m}\sim 10^3 \mathrm{cm}^2/\mathrm{Vs}$ deep in the metallic phase, which led to $n_c \sim 10^{12} \mathrm{cm}^{-2}$ and $\rho_\mathrm{c} \sim 10^3 \Omega$ \cite{wallisAndersonTransition1975, adkinsThresholdConductionInversion1978, andoElectronicPropertiesTwodimensional1982}. 
With improvement in the sample quality and substantial suppression of disorder, however, more recent ($\sim2000$) localization experiments in Si MOSFETs focused on much higher quality ($\mu_\mathrm{m} \sim 10^4 \mathrm{cm}^2/Vs$) samples, typically finding $n_c \sim 10^{11} \mathrm{cm}^{-2}$ and $\rho_c \sim 10^4 \Omega$ \cite{zavaritskayaMetalinsulatorTransitionInversion1987a, klapwijkFewElectronsIon1999, lewalleRelativeImportanceElectron2002a, tracyObservationPercolationinducedTwodimensional2009, kravchenkoPossibleMetalinsulatorTransition1994, kravchenkoScalingAnomalousMetalinsulator1995}.
The question therefore arises whether the past and the present Si MOSFET localization experiments observe the same 2D MIT phenomenon or not.

We provide in the current work a unified theory explaining both the past MOSFET experiments of the 1970s in dirty samples and the more recent MOSFET experiments in clean samples within a single theoretical framework, using a physical model of the Anderson localization crossover induced 2D MIT as arising from screened Coulomb disorder \cite{hamiltonMetallicBehaviorDilute2001, dassarmaScreeningTransport2D2015} with decreasing density leading to the strong enhancement of the effective screened disorder which causes the Anderson localization. The large quantitative differences in the values of $n_\mathrm{c}$ and $\rho_\mathrm{c}$ between the past and the present arise in our theory from the difference in the amount of random disorder (i.e., unintentional background charged impurity density in the system) in the old and the current samples, as reflected by the (at least) one order of magnitude difference in their optimal mobility.

{\em Theory.}---  
We use the well-known and extensively used Ioffe-Regel-Mott (IRM) criterion for defining the Anderson localization crossover point:
\begin{equation}
    k_\mathrm{F}l=1,
    \label{eq:IRM_criterion_kFl}
\end{equation}
where $k_\mathrm{F} = (2\pi n/g_\mathrm{v})^{1/2}$ is the 2D Fermi wave vector (where $g_\mathrm{v}$ is the possible valley degeneracy and a spin degeneracy of 2 is included) and `$l$' is the scattering mean free path given by $l= v_\mathrm{F} \tau$, where $v_\mathrm{F} = \hbar k_\mathrm{F}/m$ is the Fermi velocity and $\tau$ is the scattering time. Using the Fermi energy $E_\mathrm{F} = (\hbar^2 k_\mathrm{F}^2/2m)^{1/2}$, we can rewrite the IRM criterion as:
\begin{equation}
    E_\mathrm{F}\tau = \hbar.
    \label{eq:IRM_criterion_EfTau}
\end{equation}
The significance of the IRM criterion is that itinerant metallic electrons need to be coherent on the length scale of its wavepacket size, i.e., $1/k_\mathrm{F}$, so the minimal condition for metallic transport is the mean free path being larger than the wavepacket size defining the electron: $l > 1/k_\mathrm{F}$, which leads to $k_\mathrm{F}l=1$ defining the localization crossover point as a function of disorder (constraining the mean free path) and/or carrier density (constraining the Fermi momentum).
The Drude formula defines the resistivity $\rho$ as:
\begin{equation}
    \rho = \frac{m}{ne^2\tau}
    \label{eq:Drude}
\end{equation}
In 2D systems, $\rho$ has the dimensions of resistance and is nothing other than the resistance per square measured in ohms.  Expressing the formula above for $k_\mathrm{F}$ and  $E_\mathrm{F}$ in terms of 2D density $n$, we get from Eq.~(\ref{eq:Drude}):
\begin{equation}
    \rho= \frac{h}{e^2} \frac{k_\mathrm{F}  l} {g_\mathrm{v}}
    \label{eq:Drude2}
\end{equation}
Thus, the 2D resistivity is expressed in units of the resistance quantum, $h/e^2 = 25,812.8 \Omega$, and the nominal IRM criterion, defined by Eq.~(\ref{eq:IRM_criterion_kFl}), then gives (assuming $g_\mathrm{v}=1$ or 2 ) for the critical resistivity $\rho_\mathrm{c}$ defining the 2D MIT crossover:
\begin{equation}
  \rho_\mathrm{c} =
    \begin{cases}
      h/e^2 \sim 26 k\Omega \;\parbox{15em}{(for single valley, $g_\mathrm{v}=1$)} \\
      \mathrm{or}\\
      h/2e^2 \sim 13 k\Omega \;\parbox{15em}{(for $g_\mathrm{v}=2$ as it is for Si 100 surface MOSFETs)}
    \end{cases}       
    \label{eq:critical_resistivity}
\end{equation}
Equation~(\ref{eq:critical_resistivity}) decisively shows the theoretical conundrum we face, all Si MOSFETs,  totally independent of their disorder content,  should have exactly the same critical crossover resistivity of $h/e^2$ independent of the observed critical density $n_\mathrm{c}$, if the nominal valley degeneracy is lifted (which happens sometimes because of sharp Si-SiO$_2$ semiconductor-oxide interface breaking bulk symmetries) or $h/2e^2$, if the valley degeneracy is 2 (as it is because of the bulk valley degeneracy of 6 for the Si conduction band). Note that changing the IRM criterion by some factor, e.g., changing $\lambda_\mathrm{F}=1/k_\mathrm{F}$ to $\lambda_\mathrm{F} =2\pi/k_\mathrm{F}$, the electron wavelength, does not help because it just modifies the critical crossover resistivity by a constant factor (e.g., $2\pi$) without imparting any disorder or carrier density dependence  to the universal 2D MIT crossover resistance $\rho_\mathrm{c}$. How can this manifest disorder-independent 2D critical resistance be reconciled with the experimentally observed huge (by an order of magnitude) difference between the reported large $\rho_\mathrm{c}$ in current low-disorder samples versus the reported small $\rho_\mathrm{c}$ reported in the older high-disorder samples?

We resolve the conundrum by asserting that the IRM criterion should incorporate NOT the transport scattering time (or the transport mean free path), as is universally and uncritically assumed, but the single-particle (sometimes also called `quantum') scattering time or mean free path. The two could be different in principle because the transport quantities must include vertex corrections to the two-particle propagators whereas the single particle quantities are related to the imaginary part of single particle self-energy. The single-particle, $\tau_\mathrm{q}$, and the transport, $\tau_\mathrm{t}$, scattering times, although being formally different, often turn out to be essentially identical because vertex corrections vanish for isotropic short-range disorder potential, which is often the case. In particular, in 3D electronic materials (e.g., metals) the disorder potential is universally short-ranged and $\tau_\mathrm{t}=\tau_\mathrm{q}$ as an identity. But in 2D semiconductors, $\tau_\mathrm{t} > \tau_\mathrm{q}$ in general, and it is even possible that $\tau_\mathrm{t}\gg \tau_\mathrm{q}$ in modulation doped 2D structures where the dopant impurities are far away from the carriers \cite{dassarmaSingleparticleRelaxationTime1985,coleridgeLowfieldTransportCoefficients1989, dassarmaMobilityQualityTwodimensional2014c}.
In 2D semiconductors, such as Si MOSFET, the main disorder source are the random charged impurities (mostly in the SiO$_2$ oxide layer) which produce strongly momentum-dependent long-ranged potential with generally weak screening (because of relatively low effective carrier density and effective mass), and in addition, the charged impurities are often spatially separated from the carriers, making the disorder potential highly anisotropic and thus leading to different values of $\tau_\mathrm{t}$ and $\tau_\mathrm{q}$. The correct quantity to define the IRM criterion (when $\tau_\mathrm{t}$ and $\tau_\mathrm{q}$ are different) is obviously the single-particle scattering time (or the single-particle mean free path) since the IRM criterion is a condition on the coherence of the carriers themselves.  Obviously, coherence in metallic transport is defined by the magnitude of single particle scattering rate staying below its energy $E_\mathrm{F}$. So we now rewrite Eq.~(\ref{eq:IRM_criterion_EfTau}) as the correct IRM criterion for single-particle coherence defining the localization crossover from coherent metallic transport to Anderson localization:
\begin{equation}
E_\mathrm{F}  \tau_\mathrm{q} =\hbar.  
\label{eq:Anderson_criteria}
\end{equation}
We note that the localization criterion is defined by Eq.~(\ref{eq:Anderson_criteria}) whereas the resistivity is obviously given by Eq.~(\ref{eq:Drude}) with $\tau$ in Eq.~(\ref{eq:Drude}) being $\tau_\mathrm{t}$ by definition:
\begin{equation}
    \rho = \frac{m}{ne^2 \tau_\mathrm{t}}
    \label{eq:Drude_tr}
\end{equation}
If $\tau_\mathrm{q}$ and $\tau_\mathrm{t}$ are different, it is clear that $\rho_\mathrm{c}$ would typically be lower if the IRM criterion involves $\tau_\mathrm{q} (<\tau_\mathrm{t})$ as it should for defining coherent metallic conduction. The expressions for $\tau_\mathrm{t}$ and $\tau_\mathrm{q}$ are given by:
\begin{align} \label{eq:transport_relaxation_time}
    \frac{1}{\tau_\mathrm{t}(k)}&=\frac{2\pi}{\hbar} \int dz N_i(z) 
    \sum_{\bm k'}\left|u_{\bm k - \bm k'}(z)\right|^2 \\ \nonumber
    &\times
    (1-\cos{ \theta_{\bm k, \bm k'}})\delta(\epsilon_{\bm k}-\epsilon_{\bm k'}),
\end{align}
and
\begin{equation} \label{eq:quality_relaxation_time}
    \frac{1}{\tau_\mathrm{q}(k)}=\frac{2\pi}{\hbar} \int dz N_i(z) 
    \sum_{\bm k'}
    \left|u_{\bm k - \bm k'}(z)\right|^2
    \delta(\epsilon_{\bm k}-\epsilon_{\bm k'}),
\end{equation}
where $\epsilon_{\bm k}= \hbar^2k^2/2m$  is the usual parabolic energy dispersion with $m=0.19$ (in units of free electron mass) being the effective mass, $\theta$ is the scattering angle between the incoming and outgoing states ($\bm k$ and $\bm k'$), $N_\mathrm{i}(z)$ is the 3D distribution of charged impurities, and $u_{\bm k - \bm k'} $ is the screened Coulomb interaction between a charged disorder and an electron written as
\begin{equation}
    u_{\bm q}(z) = \frac{v_{\bm q}}{ \varepsilon(q)}e^{-qz}= \frac{2\pi e^2}{\varepsilon(q) \kappa q } e^{-qz}.
    \label{eq:Coulomb disorder}
\end{equation}
The exponential factor $e^{-qz}$ takes into account a spatial separation of $z$ between the charged impurity layer and the 2D electron layer, 
$\varepsilon(q)=1+v_{\bm q} \Pi (q)$ is the random phase approximation static dielectric function where $v_{\bm q}=2\pi e^2 / \kappa q$ is the Coulomb interaction with $\kappa$ denoting the background lattice dielectric constant. $\Pi (q)$ is the 2D static polarizability given by \cite{sternPolarizabilityTwodimensionalElectron1967}
\begin{align}     \label{eq:polar}
    \Pi(\bm q)=-\frac{m}{\pi \hbar^2}
		\left[1 - \Theta(q-2k_\mathrm{F})\frac{\sqrt{q^2- 4k^2_\mathrm{F} }}{q} \right].
\end{align}
Using Eq.~(\ref{eq:polar}), we can rewrite Eq.~(\ref{eq:Coulomb disorder}) as 
\begin{equation}
    u_{\bm q}(z) =  \frac{2\pi e^2}{\kappa (q + q_\mathrm{s}) } e^{-qz},
\end{equation}
where $q_s=q_\mathrm{TF}\left[1 - \Theta(q-2k_\mathrm{F})\sqrt{1- (2k_\mathrm{F}/q)^2 } \right] $ and $q_\mathrm{TF}=2me^2g_\mathrm{v}/\kappa \hbar^2$ is the Thosmas Fermi wavevector. Since the scattering rate at the Fermi surface is involved in the zero-temperature transport calculation [i.e., $\tau_\mathrm{t}(k_\mathrm{F})$ and $\tau_\mathrm{q}(k_\mathrm{F})$], it is easy to see that the momentum transfer $q=\left| \bm k - \bm k' \right |$ in  Eqs.~(\ref{eq:transport_relaxation_time}) and (\ref{eq:quality_relaxation_time}) is restricted to the range of $0<q<2k_\mathrm{F}$ for our calculations, leading to $q_\mathrm{s}=q_\mathrm{TF}$. Thus, the screened Coulomb disorder potential [Eq.~(\ref{eq:Coulomb disorder})] can be equivalently expressed as 
\begin{equation}
    u_{\bm q}(z) =  \frac{2\pi e^2}{\kappa (q + q_\mathrm{TF}) } e^{-qz}.
\end{equation}
Note that the $(1 – \cos\theta)$ factor in Eq.~(\ref{eq:transport_relaxation_time}), defining $\tau_\mathrm{t}$, arises from vertex corrections in the conductivity, which removes all forward scattering (i.e., $\theta \sim 0$) from the transport scattering rate because scattering in the forward direction does not contribute to the resistivity.  By contrast, the $(1-\cos \theta)$ factor is absent in the single-particle quantum scattering rate [Eq.~(\ref{eq:quality_relaxation_time})] since all scattering (including the forward direction $\theta\sim 0$) contributes to the quantum decoherence of the single-particle momentum eigenstates. When the disorder potential $u_{\bm q}$ is purely s-wave, i.e., short-ranged and thus independent of scattering momentum $q$, the $(1 – \cos \theta)$ factor drops out, leading to $\tau_\mathrm{t}=\tau_\mathrm{q}$, as happens in 3D metals, but not necessarily in 2D Si-MOSFETs.

For simplicity, and mainly  because our interest is a general understanding of how past and present 2D MIT experiments in Si MOSFETs can be reconciled within a unified theory even if they have very different values of $\rho_\mathrm{c}$ with  $\rho_\mathrm{c, present} \gg \rho_\mathrm{c, past}$, we use  a minimal 2-parameter ($n_\mathrm{i}$ and $d$) model for disorder:
\begin{equation}
    N_\mathrm{i}(z) = n_i \delta (z-d),    
    \label{eq:2params_model}
\end{equation}
where $n_\mathrm{i}$ is the 2D random quenched charged impurity density producing the scattering, which are placed at a distance ‘d’ from the 2D carriers in Si. In reality, of course, $N_\mathrm{i} (z)$ is an unknown 3D disorder throughout the Si MOSFET device, but it is known that the most resistive scattering arises mostly from the random charged impurities invariably present in the SiO$_2$ layer.  Therefore, the 2-parameter impurity model is the minimal sample-independent universal model consistent with the materials physics of all MOSFETs, past and present. We  also assume, without any loss of generality, a strict 2D approximation for the confined carriers with an effective mass $m$ and a background lattice dielectric constant $\kappa$ appropriate for the Si-SiO$_2$ system \cite{andoElectronicPropertiesTwodimensional1982}. 
\begin{figure}[!htb]
    \centering
    \includegraphics[width=\linewidth]{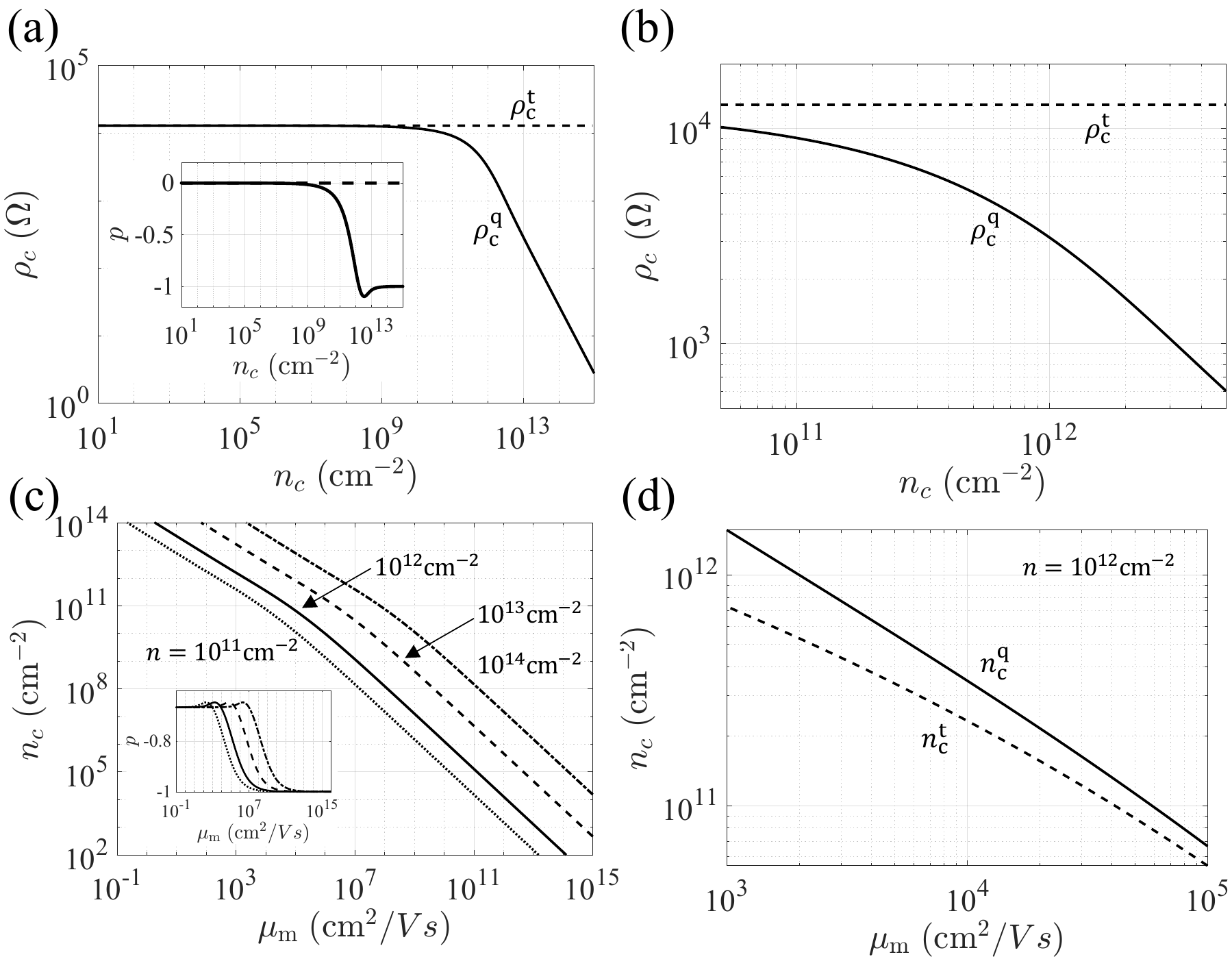}
    \caption{(a) $\rho_\mathrm{c}$ plotted as a function of $n_\mathrm{c}$ obtained using the quantum ($\rho_\mathrm{c}^\mathrm{q}$, solid line) and transport ($\rho_\mathrm{c}^\mathrm{t}$, dashed line) scattering time. The inset in (a) shows the power-law exponent numerically calculated through $p=d\ln{\rho_c}/d\ln{n_c}$. (b) Same as (a) but plotted over a typical experimental range of $n_\mathrm{c}$. (c) $n_\mathrm{c}$ (obtained using $\tau_\mathrm{q}$) plotted as a function of the mobility $\mu_\mathrm{m}$ deep in the metallic phase (i.e., $n>n_\mathrm{c}$) at carrier densities  $n=10^{11}$, $10^{12}$, $10^{13}$, and $10^{14}\mathrm{cm}^{-2}$. The inset in (c) in shows the corresponding power-law exponent $p=d\ln{n_\mathrm{c}}/d\ln{\mu_\mathrm{m}}$. (d) Plots of $n_\mathrm{c}$ obtained using $\tau_\mathrm{t}$ (dashed) and $\tau_\mathrm{q}$ (solid) at a fixed carrier density $n=10^{12} \mathrm{cm}^{-2}$ over a typical experimental range of of $\mu_\mathrm{m}$. }  
    \label{fig:1}
\end{figure}
In the next section, we present our results based on the theory above, showing the calculated $\rho_\mathrm{c}$, obtained from the modified IRM criterion using $\tau_\mathrm{q}$, as a function of the calculated critical density $n_\mathrm{c}$ (also obtained from the modified IRM criterion) as well as the calculated $n_\mathrm{c}$ as a function of the `maximum' mobility $\mu_\mathrm{m}$ deep in the metallic phase characterizing the sample quality.

{\em Results.}---  
In Fig.~\ref{fig:1} (a) and (b), we show the calculated critical resistivity obtained using $\tau_t$ (labelled as $\rho^\mathrm{t}_\mathrm{c}$) and $\tau_q$ (labelled as $\rho^\mathrm{q}_\mathrm{c}$) as a function of $n_\mathrm{c}$. Note that $\rho^t_c\sim 13 k\Omega$ is given as a constant independent of $n_\mathrm{c}$ in agreement with Eq.~(\ref{eq:critical_resistivity}). $\tau_q$ exhibits a similar flat behavior at low $n_\mathrm{c}$ but starts deviating from $\tau_t$ with a strong dependence on $n_\mathrm{c}$ at higher $n_c > 10^9 \mathrm{cm}^{-2}$, consistent with the experimental findings in Si MOSFET systems. Figure~\ref{fig:1} (c) and (d) present $n_\mathrm{c}$ as a function of the mobility $\mu_\mathrm{m}=e\tau_\mathrm{t}/m$ with $\tau_\mathrm{t}$ being calculated at high carrier densities deep in the metallic regime away from the critical density (i.e., $n>n_c$), characterizing the sample quality. In Fig.~\ref{fig:1} (c), we present several plots of $n_\mathrm{c}$ calculated at various carrier densities along with its power-law exponent plotted in the inset. It is worth noting that $n_\mathrm{c}$ decreases with increasing $\mu_\mathrm{m}$ (i.e., increasing sample quality) scaling as $n_c\sim \mu_\mathrm{m}^{-p}$ with the calculated exponent, with the calculated exponent, $0.6<p<1$, showing a weak density dependence, as observed experimentally.
Our theoretical exponent value for $p$ agrees with an empirical finding of $p$ extracted from  the recent 2D MIT experimental data \cite{sarachikDisorderdependenceCriticalDensity2002}.  Another experiment already pointed out in 1999 $\rho_\mathrm{c}$ ($n_\mathrm{c}$) increases (decreases) with decreasing sample mobility in good agreement with our results \cite{urenLogarithmicCorrectionsTwodimensional1981}.
Our results clearly show, within one unified theory, that the past and the present 2D MIT experiments in Si MOSFETs manifest identical Anderson localization physics, with the crossover critical density $n_\mathrm{c}$ (the crossover critical resistance $\rho_\mathrm{c}$) defining the transition/crossover point increasing (decreasing) with decreasing sample quality (i.e., decreasing $\mu_\mathrm{m}$). Thus, there is no new physics in the recent 2D MIT experiments for the Anderson localization crossover itself in spite of improvement in the sample mobility over the last 50 years—the transition itself is exactly the same 2D localization crossover approximately defined by the Ioffe-Regel-Mott criterion as was first observed in 1973-75.

{\em Conclusions.}---  
We have shown, using a single unified theory, that 2D MIT crossovers observed in the past and current 2D Si-MOSFET samples essentially arise from the same Anderson localization physics associated with screened Coulomb disorder. We use a realistic model for the Boltzmann transport calculations in 2D Si-MOSFET systems considering screening effects on charged impurities and the separation of the impurity layer away from the 2D electron layer. We find that the IRM criterion incorporating the single-particle scattering time $\tau_\mathrm{q}$ leads to qualitatively consistent estimates of the critical points (i.e., the critical resistivity $\rho_\mathrm{c}$ and density $n_\mathrm{c}$) with the past and current 2D MIT experiments where $n_\mathrm{c}$ ($\rho_\mathrm{c}$) decrease (increase) with increasing sample quality, whereas the IRM criterion incorporating the transport-particle scattering $\tau_\mathrm{t}$ gives sample-quality-independent universal $\rho_\mathrm{c}$, disagreeing with the experimental observations. Our work suggests that the strong localization driven by screened Coulomb disorder is primarily responsible for both past and recent 2D MIT physics in Si systems and there is no new essential physics involved in recent 2D MIT despite the vast sample quality improvement.

\section{Acknowledgement} \label{sec:acknowledgement}
This work is supported by the Laboratory for Physical Sciences.

%

\end{document}